# Two New Algorithms for Line Clipping in *E²* and Their Comparison

Václav Skala[1] and Duc Huy Bui
Department of Informatics and Computer Science[2]
The University of West Bohemia
Univerzitni 22, Box 314, 306 14 Plzeň
Czech Republic
{skala | bui} @kiv.zcu.cz       http://iason.zcu.cz/{~skala | ~bui}

**Abstract**

*New efficient algorithms for the line clipping by the given rectangle in E² are presented. The first algorithm is based on the line direction evaluation, the second one is based on a new coding technique of clipping rectangle's vertices. It allows solving all cases more effectively. A comparison of the proposed algorithms with the Liang-Barsky algorithm and some experimental results are presented, too.*

## 1. Introduction

Many algorithms for clipping a line by a rectangular area or a convex polygon in $E^2$ or by a non-convex or convex polyhedron in $E^3$ have been published so far, see [4], [5], [6], [7], [8] for main references. The line segment clipping by rectangular window in $E^2$ is often restricted to the use of the Cohen-Sutherland (CS) algorithm [1] or its modifications based on some presumptions like small clipping window or more sophisticated coding technique [9], etc. The line clipping problem solution is a bottleneck of many packages and applications and, therefore, it would be desirable to use the fastest algorithm even though it is more complex.

In many applications it is necessary to clip lines instead of line segments. It can be shown that the CS algorithm is faster than the Liang-Barsky (LB) algorithm [1] for the line segment clipping, but for the line clipping the LB algorithm is more convenient and faster.

## 2. LB Algorithm

The LB algorithm is based on clipping of the given line by each boundary line on which the rectangle edge lies. Let us assume that we have a line *p* defined by two points $A(x_A, y_A)$ and $B(x_B, y_B)$. In the LB algorithm, the given line is parametrically represented. At the beginning of computation, the parameter *t* is unlimited i.e. $t \in (-\infty, +\infty)$ and then this interval is subsequently curtailed by all the intersection points with each boundary line of the clipping rectangle, see [2], [3].

A weakness of the LB algorithm is the need to compute the parameter *t* of those intersection points, which are not part of the result. For example, see the line *p* in Figure 1, all four values *t*, representing intersection points with each boundary line, are computed but only two are valid. Some considerations how to improve the efficiency of the LB algorithm resulted into the new LSA and SF algorithms for the line clipping.

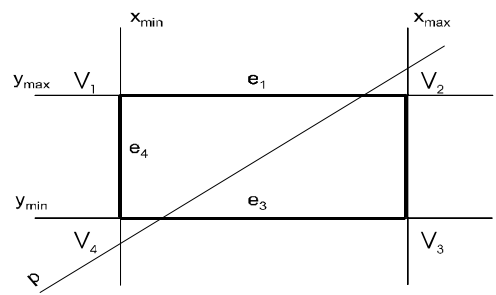

---

[1] Affiliated with the Multimedia Technology Research Centre, University of Bath, U.K.
[2] This work was supported by The Ministry of Education of the Czech Republic: project VS 97155 and project GA AV A2030801.

**Figure 1**: Line clipping against a rectangular window

## 3. Proposed methods

### 3.1. LSA Algorithm

The LSA algorithm for the line clipping is a straightforward modification of the LSSA algorithm for the line segment clipping, see [8]. The LSA algorithm is based on the simple evaluation of line direction because the usual coding scheme of the line segment's end-points cannot be used. The comparison between directions of the given line and clipping rectangle's diagonal decides which edges (horizontal or vertical) are used first to compute the intersection points between the line and the clipping window. This approach enables to avoid the calculation of the intersection points that do not lie on the boundary edges of the clipping rectangle.

### 3.2. SF Algorithm

The proposed separation function (SF) algorithm for the line clipping is based on a new coding technique for vertices of the given clipping rectangle. The given line $p$ divides the whole plane into two regions. The separation function is defined as:

$$F(x,y) = a*x + b*y + c$$

where
$a = \Delta y = y_B - y_A$,
$b = -\Delta x = x_A - x_B$,
$c = x_B*y_A - x_A*y_B$.

The sign of the separation function value $F(V_i)$ ($i \in [1,4]$) in the $i$-th vertex of the rectangular window determines the region in which the vertex lies. Using the value of the separation function in all vertices we can distinguish 7 fundamental cases, see Figure 2.

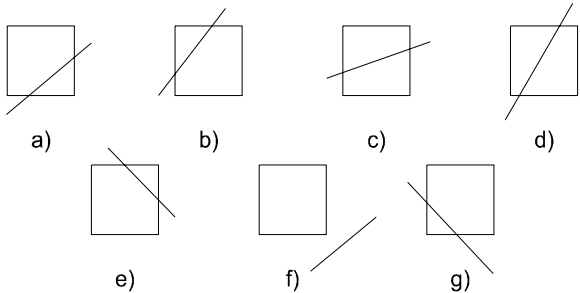

**Figure 2**: Line categorization by clipping edges

This analysis led naturally to the SF algorithm. The basic steps can be defined as:

- calculate the coefficients *a, b, c*,
- use the separation function $F$ to characterise the location of vertices of the given clipping rectangle,
- determine the appropriate case,
- calculate the intersection points with appropriate edges.

It can be seen that only the intersection points required for the output are computed.

We will describe now the classification process more in detail. Let us denote $c_1$, $c_2$, $c_3$, $c_4$ values of the separation function in the clipping rectangle's vertices $V_1$, $V_2$, $V_3$, $V_4$, see Figure 1.

There are two major cases to be distinguished:

- the vertices $V_1$ and $V_3$ lie on the different sides of line $p$, see Figure 2.a-d,
- the vertices $V_1$ and $V_3$ lie on the same side of line $p$, see Figure 2.e-f.

**a) The vertices $V_1$ and $V_3$ are in the different sides of the line, i.e. $c_1 * c_3 < 0$**. In this case, the sign of expression $(c_2*c_4)$ determine whether $V_2$ and $V_4$ lie on the same sides of line $p$ (Figure 2.a-b) or not (Figure 2.c-d).

If the vertices $V_2$ and $V_4$ lie on the same side of line $p$, the additional test $c_1*c_2>0$ determines on which edge the intersection points lie. If $c_1*c_2 > 0$ then the intersection points lie on the edges $e_2$ and $e_3$, see Figure 2.a. Otherwise, the intersection points lie on the edges $e_1$ and $e_4$, see Figure 2.b.

In the case, when $V_2$ and $V_4$ lie on the different sides of line $p$, if $c_1*c_2 > 0$ then the intersection points lie on the edges $e_2$ and $e_4$, see Figure 2.c. Otherwise, the intersection points lie on the edges $e_1$ and $e_3$, see Figure 2.d.

**b) The vertices $V_1$ and $V_3$ lie on the same side of the line**. In this case, if $c_1*c_2 < 0$, i.e. $V_1$ and $V_2$ lie on the different sides of the line, then the intersection points lie on the edges $e_1$ and $e_2$, see Figure 2.e. Otherwise, the additional test $c_1*c_4 > 0$ determines whether the whole line is outside of clipping rectangle, see Figure 2.f, or the intersection points lie on the edges $e_3$ and $e_4$, see Figure 2.g.

The complete SF algorithm can be implemented by the Algorithm 1.

### 3.3 Modified SF algorithm

It can be seen that some modifications can be done to improve the efficiency of SF algorithm.

a) The first modification is based on the observation that the co-ordinates of the intersection points can be calculated from the separation function value of the clipping window's vertices. It is very simple to derive the following expressions:

- $x = x_A + (y_{max} - y_A) * \Delta x/\Delta y = x_{min} - c_1/\Delta y$
  (the intersection point on the top boundary)

- $y = y_A + (x_{max} - x_A) * \Delta y/\Delta x = y_{max} + c_2/\Delta x$
  (the intersection point on the right boundary)

- $x = x_A + (y_{min} - y_A) * \Delta x/\Delta y = x_{min} - c_4/\Delta y$
  (the intersection point on bottom boundary)

- $y = y_A + (x_{min} - x_A) * \Delta y/\Delta x = y_{max} + c_1/\Delta x$
  (the intersection point on the left boundary)

It can be seen that we can save one addition and one multiplication for each intersection point by using these expressions.

b) Better results can be obtained while replacing the direct calculation of $c_2$, $c_3$, $c_4$ by using the pre-calculated values as follows:

- $c_2 = \Delta y * x_{max} - \Delta x * y_{max} + c = c_1 + \Delta y * w$
- $c_3 = \Delta y * x_{max} - \Delta x * y_{min} + c = c_2 + \Delta x * h$
- $c_4 = \Delta y * x_{min} - \Delta x * y_{min} + c = c_1 + \Delta x * h$

where:
$w = x_{max} - x_{min}$ (the clipping window's width)
$h = y_{max} - y_{min}$ (the clipping window's height)

c) We can get further speed-up when applying the following replacements: instead of two statements ($c := x_B * y_A - x_A * y_B$; $c_1 := \Delta y * x_{min} - \Delta x * y_{max} + c$), we can use only one ($c_1 := \Delta y * (x_{min} - x_A) - \Delta x * (y_{max} - y_A)$) and instead of the condition ($c_1 * c_3 < 0$), we can use the condition ($c_1 * (c_2 + \Delta x * h) < 0$).

All of above mentioned modifications can be implemented by the MSF algorithm, see Algorithm 2.

Since the conditions $\Delta x = 0$, $\Delta y = 0$ occur practically with **zero probability**, the test of these conditions can be left out and the further speed-up can be reached. This algorithm will be reported as the **MSF-1** algorithm.

### 4. Experimental results

For experimental verification of the LB and LSA algorithms and the proposed SF, MSF and MSF-1 algorithms, all fundamental cases were tested and $8.10^6$ different lines were randomly generated for each considered case, see Figure 3 and Figure 4.

Let us introduce the coefficients of efficiency $v_{LSA}$, $v_{MSF}$, $v_{MSF-1}$ as:

$$v_{LSA} = \frac{T_{LB}}{T_{LSA}}, \quad v_{MSF} = \frac{T_{LB}}{T_{MSF}}, \quad v_{MSF-1} = \frac{T_{LB}}{T_{MSF-1}}$$

where $T_{LB}$, $T_{LSA}$, $T_{MSF}$, $T_{MSF-1}$ are times consumed by the LB, LSA algorithm and the modifications MSF and MSF-1 algorithms of the SF algorithm (the MSF-1 algorithm is the MSF algorithm without testing of conditions $\Delta x = 0$, $\Delta y = 0$).

Table 1 contains experimental results obtained for Pentium II-350MHz/64MB RAM/512KB CACHE, similar results were also obtained for Pentium-75MHz/32MB RAM and Pentium PRO-200MHz/128MB RAM. This table shows that the LSA and MSF algorithms are significantly faster than the LB algorithm in all cases. It can be seen that the speed-up varies from 1.3 to 1.87 for all common cases. The common case is the case when the given line is neither horizontal nor vertical (cases $p_1$-$p_7$).

The MSF-1 algorithm is based on the fact that strictly horizontal or vertical lines are highly improbable in normal situations. The speed-up of the MSF-1 algorithm can be expected from 1.7 to 2.16 for all common cases, see Table 1.

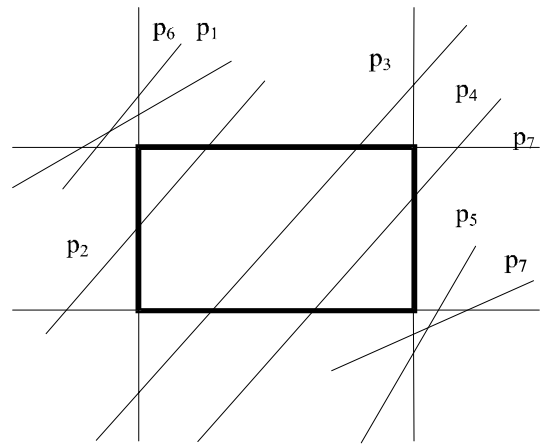

**Figure 3**: Generic lines for algorithms comparison

```
procedure SF_Clip ( xA, yA, xB, yB: real); {points A(xA, yA,) and B(xB, yB) determine the clipped line}
global var xmin, xmax, ymin, ymax: real;  { co-ordinates of clipping window corners }
var     t, Δx, Δy, c, c1, c2, c3, c4 : real;
begin   Δx := xB - xA;
        if  Δx = 0 then
                begin   if (xA < xmin) or (xA > xmax) then EXIT; {the line is outside of the rectangle}
                        yA := ymin; yB := ymax;
                        DRAW_LINE (xA, yA, xB, yB); EXIT {SF_Clip}
                end;
        Δy := yB - yA;
        if  Δy = 0 then
                begin   if (yA < ymin) or (yA > ymax) then EXIT; {the line is outside of the rectangle}
                        xA := xmin; xB := xmax;
                        DRAW_LINE (xA, yA, xB, yB); EXIT {SF_Clip}
                end;
        c := xB * yA − xA * yB;        c1:= Δy * xmin − Δx * ymax + c;
        c2:= Δy * xmax − Δx * ymax + c;    c3:= Δy * xmax − Δx * ymin + c;
        if (c1 * c3 < 0) then
                begin  c4:= Δy * xmin − Δx * ymin + c;
                       if (c2 * c4 > 0) then
                                if (c1 * c2 > 0) then                      {case a}
                                        begin   yB := yA + (xmax - xA) * Δy / Δx; xB := xmax;
                                                xA := xA + (ymin − yA) * Δx / Δy; yA := ymin
                                        end
                                else                                       {case b}
                                        begin   xB := xA + (ymax − yA) * Δx / Δy; yB := ymax;
                                                yA := yA + (xmin - xA) * Δy / Δx; xA := xmin
                                        end
                       else                                                {(c2 * c4 < 0)}
                                if (c1 * c2 > 0) then                      {case c}
                                        begin    t := Δy / Δx;
                                                 yB := yA + (xmax - xA) * t; xB := xmax;
                                                 yA := yA + (xmin - xA) * t; xA := xmin
                                        end
                                else    begin    t := Δx / Δy;    {case d}
                                                 xB := xA + (ymax − yA) * t; yB := ymax;
                                                 xA := xA + (ymin − yA) * t; yA := ymin
                                        end
                end
        else                 { (c1 * c3 > 0) }
                begin   if (c1 * c2 < 0) then                              {case e}
                                begin   yB := yA + (xmax - xA) * Δy / Δx; xB := xmax;
                                        xA := xA + (ymax − yA) * Δx / Δy; yA := ymax
                                end
                        else     begin   c4:= Δy * xmin − Δx * ymin + c;
                                         if (c1 * c4 > 0) then EXIT;       {case f}
                                         else                              {case g}
                                                begin   xB := xA + (ymin − yA) * Δx / Δy; yB := ymin;
                                                        yA := yA + (xmin - xA) * Δy / Δx; xA := xmin
                                                end
                                 end
                end;
        DRAW_LINE (xA, yA, xB, yB)
end { SF_Clip };
```

**Algorithm 1**: SF algorithm

```
procedure MSF_Clip ( xA, yA, xB, yB: real); {points A(xA, yA,) and B(xB, yB) determine the clipped line}
global var xmin, xmax, ymin, ymax, h, w: real; {corners' co-ordinates and size of clipping window}
var     t, Δx, Δy, c1, c2, c3, c4 : real;
begin   Δx := xB - xA;
        /* {tests if line is vertical or horizontal}
        if  Δx = 0 then
                begin   if (xA < xmin) or (xA > xmax) then EXIT; {the line is outside of the rectangle}
                        yA := ymin; yB := ymax;
                        DRAW_LINE (xA, yA, xB, yB); EXIT {MSF_Clip}
                end;
        Δy := yB - yA;
        if  Δy = 0 then
                begin   if (yA < ymin) or (yA > ymax) then EXIT; {the line is outside of the rectangle}
                        xA := xmin; xB := xmax;
                        DRAW_LINE (xA, yA, xB, yB); EXIT {MSF_Clip}
                end;
        /*{end of the section to be removed for MSF-1 algorithm}
        c1:= Δy * (xmin – xA) - Δx * (ymax - yA);        c2:= c1 + Δy * w;
        if c1 * (c2 + Δx * h) < 0 then
                begin   c4:= c1 + Δx * h;
                        if (c2 * c4 > 0) then
                                if (c1 * c2 > 0) then                           {case a}
                                        begin   yB := ymax + c2 / Δx; xB := xmax;
                                                xA := xmin – c4 / Δy; yA := ymin
                                        end
                                else                                             {case b}
                                        begin   xB := xmin – c1 / Δy; yB := ymax;
                                                yA := ymax + c1 / Δx; xA := xmin
                                        end
                        else                                                    { (c2 * c4 < 0) }
                                if (c1 * c2 > 0) then                           {case c}
                                        begin   t := 1.0 / Δx;
                                                yB := ymax + c2 * t;    xB := xmax;
                                                yA := ymax + c1 * t;    xA := xmin
                                        end
                                else    begin   t := 1.0 / Δy;    {case d}
                                                xB := xmin – c1 * t;    yB := ymax;
                                                xA := xmin – c4 * t;    yA := ymin
                                        end
                end
        else            { (c1 * c3 > 0) }
                begin   if (c1 * c2 < 0) then                                   {case e}
                                begin   yB := ymax + c2 / Δx; xB := xmax;
                                        xA := xmin – c1 / Δy; yA := ymax
                                end
                        else    begin   c4:= c1 + Δx * h;
                                        if (c1 * c4 > 0) then EXIT;             {case f}
                                        else                                    {case g}
                                                begin   xB := xmin – c4 / Δy;        yB := ymin;
                                                        yA := ymax + c1 / Δx; xA := xmin
                                                end
                                end
                end;
        DRAW_LINE (xA, yA, xB, yB)
end { MSF_Clip };
```

**Algorithm 2**: MSF algorithm

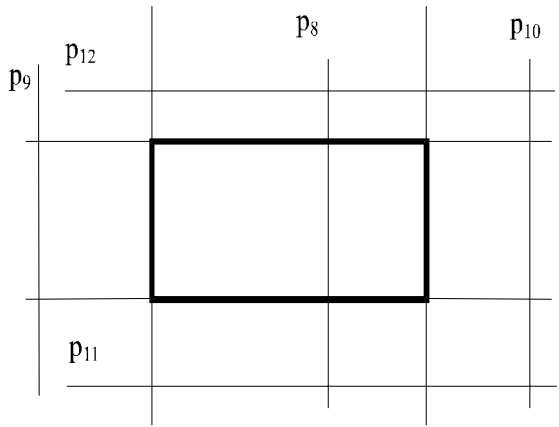

**Figure 4**: Horizontal and vertical lines

|  | Pentium II 350MHz/64MB RAM | | |
|---|---|---|---|
| **Case** | $v_{LSA}$ | $v_{MSF}$ | $v_{MSF-1}$ |
| $P_1$ | 1.63 | 1.87 | 2.16 |
| $P_2$ | 1.55 | 1.59 | 1.70 |
| $P_3$ | 1.65 | 1.74 | 1.87 |
| $P_4$ | 1.30 | 1.53 | 1.68 |
| $P_5$ | 1.42 | 1.49 | 1.72 |
| $P_6$ | 1.81 | 1.87 | 2.16 |
| $P_7$ | 1.37 | 1.49 | 1.72 |
| $P_8$ | 2.62 | 2.62 | 1.52 |
| $P_9$ | 1.16 | 1.16 | 0.75 |
| $p_{10}$ | 1.26 | 1.26 | 0.94 |
| $p_{11}$ | 2.19 | 2.20 | 1.55 |
| $p_{12}$ | 2.17 | 2.19 | 1.75 |

**Table 1**: Experimental results

## 5. Conclusion Acknowledgements

The both new LSA and SF (including its modifications) algorithms for line clipping against a given rectangle in $E^2$ were developed, verified and tested. The proposed algorithms are convenient for all applications, especially when many lines must be clipped. These algorithms give similar efficiency but the LSA algorithm is simpler to implement and interpret. The LSA and MSF algorithm claim the superiority over the LB algorithm for all cases. Experiments proved that the speed-up can be considered up to 1.6 times on the average for the LSA algorithm. The MSF and MSF-1 algorithms were also implemented as the modifications of the SF algorithm. The speed-up of MSF-1 algorithm can be considered up to **1.86** times on the average for all common cases.

The new developed algorithms proved that the approach "test first and compute after all tests" can bring a significant speed-up. There is a hope, that the LSA algorithm and modifications of the SF algorithm can be extended to $E^3$ case and implemented in hardware, too.

Some related reports are available in the on-line form at the URL:

http://herakles.zcu.cz/publication.htm

**Acknowledgements**

The authors would like to express their thanks to all who contributed to this work, mainly to recent MSc. and PhD. students at the University of West Bohemia in Plzen, who stimulated this work and especially to Dr. Ivana Kolingerova for extremely constructive criticisms.